\documentclass[aps,pra,twocolumn]{revtex4}
\usepackage{eurosym}
\usepackage{graphics}
\usepackage[thmmarks]{ntheorem}
\usepackage{graphicx}
\usepackage{amsmath}
\usepackage{amsfonts}
\usepackage{amssymb}
\usepackage{float}
\usepackage{longtable}
\usepackage{epsfig}
\usepackage{latexsym}
\usepackage{theorem}
\usepackage{bbm}
\usepackage{bm}
\usepackage{psfrag}
\usepackage{color}

\newcommand{\PP}[1]{{{\textcolor{black}{#1}}}}
\usepackage{xcolor}
\usepackage{hyperref}

\DeclareMathOperator{\diag}{diag}

\begin{document}

\title{Continuous-variable quantum cryptography with discrete alphabets:\\
Composable security under collective Gaussian attacks}
\author{Panagiotis Papanastasiou}
\author{Stefano Pirandola}
\affiliation{Department of Computer Science, University of York, York YO10 5GH, United
Kingdom}

\begin{abstract}
We consider continuous-variable quantum key distribution with
discrete-alphabet encodings. In particular, we study protocols where
information is encoded in the phase of displaced coherent (or thermal)
states, even though the results can be directly extended to any protocol
based on finite constellations of displaced Gaussian states. In
this setting, we study the composable security in the finite-size
regime assuming the realistic but restrictive hypothesis of collective
Gaussian attacks. Under this assumption, we can efficiently estimate the
parameters of the channel via maximum likelihood estimators and bound the
corresponding error in the final secret key rate.
\end{abstract}

\maketitle

\section{Introduction}
Quantum key distribution (QKD)~\cite{bb84,E91,BBM92,Stefano_rev2} allows two
remote authenticated parties to establish a shared secret key without any
assumption on the computational power of the eavesdropper, the security
being based on fundamental laws of quantum mechanics, such as the no-cloning
theorem~\cite{noclone,noclone0}. The first QKD\ protocols were based on the
use of discrete variables (DVs), i.e., discrete degrees of freedom of the
electromagnetic field, such as polarization or time bins. Later, at the end
of the 90s and beginning of 2000, QKD was extended to
continuous-variables (CVs)~\cite{Sam_rev,Stefano_rev} by the work of Ralph~%
\cite{CV1} and other authors~\cite{CV2,CV3,CV4,CV5}, culminating in the
seminal GG02 protocol~\cite{GG02} based on Gaussian modulation of coherent
states. This seminal work also introduced the notion of reverse
reconciliation that allowed experimental CV-QKD to reach long distances and
led to the theoretical introduction of the reverse coherent information of a
bosonic channel~\cite{rev1,rev2}. The authors of Ref.~\cite{rev2}  first explored
 the ultimate limits of point-to-point (i.e., repeaterless) QKD,
culminating in 2015 with the discovery and proof of the Pirandola-Laurenza-Ottaviani-Banchi (PLOB) bound~\cite%
{PLOB} (see Refs.~\cite{Stefano_rev2,TQC} for more details on the historical
developments).

Other theoretical advances in CV-QKD were the introduction of thermal-state
protocols~\cite{thermal1,thermal2,thermal3,thermal4,usenkotrusted,panos_thermal} (where the authors of
Refs.~\cite{thermal3,thermal4} specifically studied the extension to longer
wavelengths, down to the microwaves), two-way quantum communication
protocols~\cite{two-way,Carlo_two-way}, one-dimensional protocols~\cite%
{1dim,1dim2}, and CV\ measurement-device independent (MDI) QKD~\cite%
{stefano_mdi}. Last but not least, there was the important development of
CV-QKD\ with discrete-alphabet encoding. This idea was first introduced in
the post-selection protocol of the authors of Ref.~\cite{disc1} and later developed in a
number of works~\cite{disc2,disc3,disc4,disc5,disc6,disc7,disc8,disc9,disc10}.
In particular, the authors of Refs.~\cite{disc1,disc2,disc3,disc4} considered binary and
ternary alphabets of displaced coherent states. The authors of Ref.~\cite{disc7} considered
four coherent states and, later, other works studied alphabets with
arbitrary number of states under pure-loss~\cite{disc5} and thermal-loss~%
\cite{disc6} attacks. All these security proofs were limited to the
asymptotic case of infinite signals exchanged by the parties. In particular,
the security of discrete-alphabet CV\ QKD has been proven asymptotically
under  collective attacks using decoy-like states in Ref.~\cite{disc8}  and, 
more recently, under general attacks using a Gaussian bound in Ref.~\cite{disc9} 
(see also Ref.~\cite{disc10}).

In this work, we depart from the asymptotic security assumption and study the finite-size composable security of
discrete-alphabet CV-QKD\ protocols. However, this extension comes with the price of another restriction. In fact, our analysis holds under the assumption of collective
Gaussian attacks~\cite{collectiveG} and, in particular, collective
entangling cloner attacks~\cite{virtual,Stefano_rev2} which results into a
realistic thermal-loss channel between the remote parties. While the
general arguments apply to any discrete alphabet, we focus on the case of
phase-encoded coherent (or thermal) states, so that they are displaced in
the phase-space to create regular constellations at fixed distance from the
vacuum state. Our techniques combine tools from Refs.~\cite{Antony_cpe,Usenko_cpe,panos_thermal,compLeverier1,Cosmo_comp,TomaThesis,Portmann,AEP,TomaRenner,free_space}. The assumption of
collective Gaussian attack is particularly useful for the purpose of parameter estimation,
for which we follow the approach found in  Refs.~\cite{Antony_cpe,Usenko_cpe,panos_thermal}.
\PP{The composable analysis resorts to various tools as combined in the approach found in Ref.~\cite{free_space} here specified and applied to a discrete-alphabet protocol.}

The paper is organized as follows. In Sec.~\ref{sec:Signal state
exchange}, we describe the discrete-alphabet (phase-encoded) QKD protocol,
for which we discuss the asymptotic security analysis. In Sec.~\ref{sec:CPE}%
, we discuss parameter estimation in the presence of finite-size effects
and, in Sec.~\ref{composableSEC},\ we study the key rate of the protocol
in the composable security framework. Section~\ref{ConclusionSEC}\ is for
conclusions.

\section{Asymptotic security of a phase-encoded protocol\label{sec:Signal
state exchange}}
In a generic phase-encoded CV-QKD protocol with $N$ states, Alice randomly
chooses between $N$ coherent states $\left\vert\alpha_k\right\rangle $ with amplitude $\alpha_k :=2^{-1} \alpha \exp (\mathrm{i}2k\pi
N^{-1})$, where $\alpha \geq 0$ and $k=0,\dots ,N-1$ (so that
the classical label $k$ is chosen with probability $P_{k}=N^{-1}$). More
generally, she prepares her mode $A$ in one of $N$ displaced thermal states $%
\rho _{A|k}$ with amplitudes $\alpha _{k}$, each with a fixed mean number of photons $\bar{n}_{\text{th}}$. In
terms of quadrature operators $\mathbf{\hat{x}}_{A}:=(\hat{q}_{A},\hat{p}%
_{A})^{T}$ (with the quantum shot noise equal to $1$), Alice's conditional
thermal state has mean value
\begin{equation}
\bar{\mathbf{x}}_{A|k}:=\mathrm{Tr}(\mathbf{\hat{x}}_{A}\rho _{k})=\alpha
\begin{pmatrix}
\cos \left( 2k\pi N^{-1}\right) \\
\sin \left( 2k\pi N^{-1}\right)
\end{pmatrix},
\end{equation}%
and covariance matrix (CM) $\mathbf{V}_{A|k}=(\nu _{\text{th}}+1)\mathbf{I}$%
, where $\nu _{\text{th}}=2\bar{n}_{\text{th}}$ and $\mathbf{I}$ is the
bidimensional identity matrix.

The signal state $\rho _{A|k}$ is traveling through a Gaussian
(thermal-loss) channel which is under the full control of Eve. This is
described by transmissivity $\tau $ and injected thermal noise $\omega \geq
1 $. This channel can always be dilated into an entangling cloner attack~%
\cite{collectiveG}, where Eve has a two-mode squeezed-vacuum (TMSV) state $%
\rho _{eE_0}$ with zero mean $\bar{\mathbf{x}}_{eE_0}=(0,0,0,0)$ and CM
\begin{equation}
\mathbf{V}_{eE_0}=%
\begin{pmatrix}
\omega \mathbf{I} & \sqrt{\omega ^{2}-1}\mathbf{Z} \\
\sqrt{\omega ^{2}-1}\mathbf{Z} & \omega \mathbf{I}%
\end{pmatrix}%
,
\end{equation}%
where $\mathbf{Z}=\diag\{1,-1\}$.
In particular, mode $e$ is mixed with Alice's traveling mode $A$ in a
beam-splitter with transmissivity $\tau $ described by the symplectic matrix%
\begin{equation}
\mathcal{B}(\tau )=%
\begin{pmatrix}
\sqrt{\tau }\mathbf{I} & \sqrt{1-\tau }\mathbf{I} \\
-\sqrt{1-\tau }\mathbf{I} & \sqrt{\tau }\mathbf{I}%
\end{pmatrix}.%
\end{equation}%
After the interaction, modes $e^{\prime }$ and $E_0$ are kept in a quantum
memory for an optimal final measurement taking into consideration all the
classical communication between the parties. For each use of the channel,
Eve's and Bob's conditional output state $\rho _{Be^{\prime }E_0|k}$ has mean value
and CM given by
\begin{align}
\bar{\mathbf{x}}_{Be^{\prime }E_0|k}& =[\mathcal{B}(\tau )\oplus \mathbf{I]}(%
\bar{\mathbf{x}}_{A|k}\oplus \bar{\mathbf{x}}_{eE_0})=\bar{\mathbf{x}}%
_{B|k}\oplus \bar{\mathbf{x}}_{e^{\prime }E_0|k}, \\
\mathbf{V}_{Be^{\prime }E_0|k}& =[\mathcal{B}(\tau )\oplus \mathbf{I}]\left(
\mathbf{V}_{A|k}\oplus \mathbf{V}_{eE_0}\right) [\mathcal{B}(\tau )^{\mathsf{T}}\oplus
\mathbf{I}]  \notag \\
& =%
\begin{pmatrix}
\mathbf{B} & \mathbf{C} \\
\mathbf{C}^{\mathsf{T}} & \mathbf{V}_{e^{\prime }E_0|k}%
\end{pmatrix}%
.
\end{align}

At the output, assume that Bob applies heterodyne measurement with outcome $%
(q_{B},p_{B})$. Then, Eve's \PP{triply} conditional state $%
\rho _{\PP{e^{\prime }E_0}|kq_{B}p_{B}}$ has mean value and CM~\cite{Cirac,Jens,stefano_gd}
\begin{align}
\bar{\mathbf{x}}_{e'E_0|q_Bp_Bk}&=\bar{\mathbf{x}}_{e'E_0|k}
-\mathbf{C}^{\mathsf{T}}(\mathbf{B}+\mathbf{I})^{-1}\left[\bar{\mathbf{x}}_{B|k}-\begin{pmatrix}q_B\\p_B\end{pmatrix}\right],\label{eq:mean heterodyne}\\
\mathbf{V}_{e^{\prime }E_0|q_{B}p_{B}k}&=\mathbf{V}_{e^{\prime }E_0|k}-\mathbf{C}^{%
\mathsf{T}}(\mathbf{B}+\mathbf{I})^{-1}\mathbf{C},  \label{eq:CM heterodyne}
\end{align}%
while the probability of the outcome is given by
\begin{equation}
P_{q_{B}p_{B}|k}=\frac{e^{-\frac{1}{2}\frac{[q_{B}-\sqrt{\tau }\alpha \cos
(2k\pi N^{-1})]^{2}+[p_{B}-\sqrt{\tau }\alpha \sin (2k\pi N^{-1})]^{2}}{%
\Omega }}}{2\pi \Omega },
\end{equation}%
with $\Omega :=2+\tau \nu _{\text{th}}+(1-\tau )(\omega -1)$. Setting
\begin{equation}
q_{B}+\mathrm{i}p_{B}=\beta e^{\mathrm{i}(2l\pi N^{-1}+\theta )},
\label{setting}
\end{equation}%
with $\beta \geq 0$ and $\theta \in \lbrack -\pi N^{-1},\pi N^{-1}]$, we
obtain
\begin{eqnarray}
P_{\beta \theta l|k} &=&\frac{1}{2\pi \Omega }e^{\frac{-[\beta \cos (2l\pi
N^{-1}+\theta )-\sqrt{\tau }\alpha \cos (2l\pi N^{-1})]^{2}}{2\Omega }}
\notag \\
&&\times e^{\frac{-[\beta \sin \left( 2l\pi N^{-1}+\theta \right) -\sqrt{%
\tau }\alpha \sin \left( 2l\pi N^{-1}\right) ]^{2}}{2\Omega }}.
\end{eqnarray}%
Integrating over for $\beta $ and for $\theta $, we derive
\begin{equation}
P_{l|k}=\iint_{0,-\pi N^{-1}}^{\infty ,\pi N^{-1}}\beta P_{\beta \theta
l|k}d\beta d\theta ,
\end{equation}%
which can be calculated numerically. Here $l$ is Bob's estimator of Alice's
encoding variable $k$. Using Bayes' formula we may write
\begin{equation}
P_{k|l}=\frac{P_{l|k}P_{k}}{\sum_{k=0}^{N-1}P_{k}P_{l|k}},
\label{eq:phase-shift Pk|l}
\end{equation}%
and compute the residual entropy
\begin{equation}
H(k|l)=\sum_{l}P_{l}\sum_{k}\left( -P_{k|l}\log _{2}P_{k|l}\right) .
\end{equation}

\begin{figure}
\vspace{-0.2cm}
\includegraphics[width=0.42\textwidth]{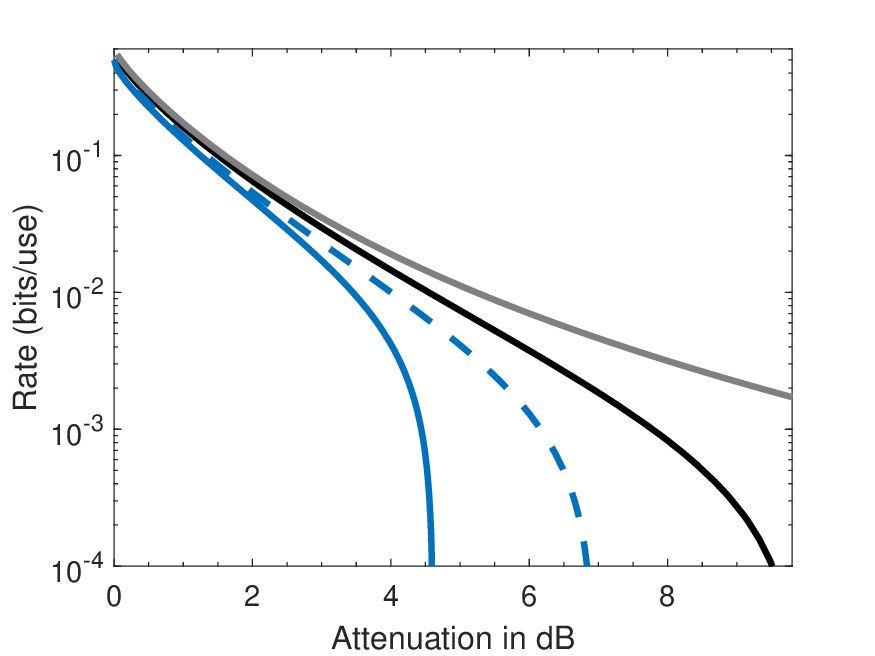}
\caption{\label{fig:plot_correct}Secret key rate for $N=2$ versus attenuation in dB. We assume $\alpha=2$ and excess noise $\varepsilon=0.01$. We show the asymptotic case with $\xi=1$  (black solid line) and the composable case, for which we assume $\epsilon_\PP{\text{s}}=\epsilon_\PP{\text{h}}=10^{-10}$, $\epsilon_\PP{\text{PE}}=10^{-10}$, $p=0.9$, $\xi=0.99$ and $r=0.01$, for $M=10^{12}$ (blue dashed line) and $M=10^{9}$ (blue solid line). All the lines have a truncation accuracy of $10$ Fock-basis states. For comparison, we also plot the corresponding asymptotic rate (grey solid line) assuming a pure loss channel.}
\end{figure}

\begin{figure}
\includegraphics[width=0.42\textwidth]{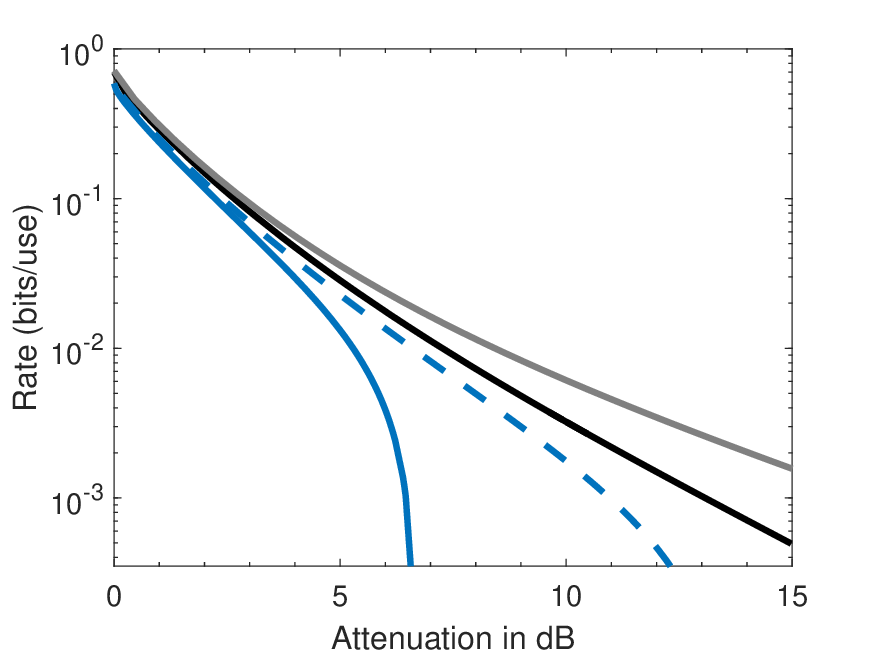}
\caption{\label{fig:plot_correctN=3}The secret key rate for $N=3$ versus attenuation in dB. We assume $\alpha=2$ and excess noise $\varepsilon=0.01$. We include the asymptotic case  with $\xi=1$ (black solid line) and the composable case, for which we assume $\epsilon_\PP{\text{s}}=\epsilon_\PP{\text{h}}=10^{-10}$, $\epsilon_\PP{\text{PE}}=10^{-10}$, $p=0.9$, $\xi=0.99$ and $r=0.01$, for $M=10^{12}$ (blue dashed line) and  $M=10^{9}$ (blue solid line). All the lines have a truncation accuracy of $10$ Fock-basis states. For comparison, we also plot the corresponding  asymptotic  rate (grey solid line) assuming a pure loss channel.}
\end{figure}

The mutual information between the variables $k$ and $l$ is given by
\begin{equation}
I(k:l)=H(k)-H(k|l)=\log _{2}N-H(k|l).  \label{eq:mi}
\end{equation}%
In reverse reconciliation (RR), Eve's information on $l$ is bounded by the
Holevo quantity
\begin{equation}
\chi (E:l)=S(\rho _{E})-\sum_{l}P_{l}S(\rho
_{E|l})  \label{holq}
\end{equation}%
with $E:=e^\prime E_0$, where $\rho _{E}:=\sum_{l}P_{l}\rho _{E|l}$ is
non-Gaussian, and the conditional state $\rho _{E|l}$ is
calculated by using the replacement of Eq.~(\ref{setting}) in the Gaussian
state $\rho _{E|q_{B}p_{B}k}$~\cite{Note} and averaging over the probability $%
P_{k\beta \theta |l}$, i.e., we have%
\begin{equation}
\rho _{E|l}=\sum_{k=0}^{N-1}\iint_{0,-\pi N^{-1}}^{\infty ,\pi
N^{-1}}P_{\beta \theta k|l}\rho _{E|\beta \theta lk}d\theta
d\beta,  \label{connd}
\end{equation}%
where
\begin{equation}
P_{\beta \theta k|l}=\frac{P_{\beta \theta l|k}P_k}{P_{l}}.
\end{equation}
Thus, the asymptotic secret key rate in RR is given by~\cite{Devetak}
\begin{equation}
R=\xi I(k:l)-\chi (E:l),  \label{eq:asymptotic rate 2}
\end{equation}%
where $\xi \in [0,1]$ is the reconciliation efficiency. In Fig.~\ref{fig:plot_correct}, we  plotted this
rate (solid black line) for the case of two states ($N=2$) with $\xi=1$, assuming
excess noise $\varepsilon:=\tau^{-1}(1-\tau)(\omega-1)=0.01$ and setting
$\alpha=2$. In Fig.~\ref{fig:plot_correctN=3}, we  showed the corresponding rate for $N=3$, assuming the same parameters.
\section{Channel parameter estimation\label{sec:CPE}}
The asymptotic rate in Eq.~(\ref{eq:asymptotic rate 2}) is a function of
Alice's encoding parameters, i.e., $\alpha $, $N$ and $\nu _{\text{th}}$,
together with the channel parameters, i.e., $\tau $ and $\omega $, or
equivalently $\tau $ and $\varepsilon $. To estimate the
parameters of the channel, Alice and Bob sacrifice $m$ signal states. By communicating their outcomes
for these $m$ signals, Alice and Bob can compute estimators for $\tau$ and $%
V_{\varepsilon }:=\tau \varepsilon $, and corresponding confidence intervals.
They can choose worst-case parameters to be used in the computation of the
key rate in Eq.~(\ref{eq:asymptotic rate 2}).

Therefore, assume that Alice reveals the encoding $k$ of $m$ signal states
out of a block of $M=m+n$ signal states. For $m$ sufficiently large, we have
that $m/N$ can be chosen to be an integer. Bob will have samples ${B_{k}}%
_{i} $ for $i=1,\dots, m/N$ associated to a specific Alice's encoding $k$.
Because we assume heterodyne detection, the discussion of the $\hat{q}$ and $%
\hat{p}$ quadratures is symmetric. In the $\hat{q}$ quadrature, Bob's
sampled $q$-quadratures ${B_{k}}_{i}$ can be described by the following
stochastic variable
\begin{align}
q_{B_{k}}& =\sqrt{\frac{\tau }{2}}\alpha \cos \left( 2k\pi /N\right) +q_{%
\text{no}},\label{eq:arriving signal} \\
q_{\text{no}}& :=\sqrt{\frac{\tau }{2}}q_{\text{th}}+\sqrt{\frac{1-\tau }{2}}%
q_{E}+\sqrt{\frac{1}{2}}q_{\text{h}},
\end{align}%
where $q_{\text{th}}$ is Alice preparation noise with variance $\nu _{\text{%
th}}+1$, $q_{E}$ is Eve's noise variable with variance $\omega $, and $q_{%
\text{h}}$ is the noise variable due to Bob's heterodyne measurement. The
variable $q_{B_{k}}$ is Gaussian with mean
\begin{equation}
\mathbb{E}(q_{B_{k}})=\sqrt{\frac{\tau }{2}}\alpha \cos \left( 2k\pi
/N\right),
\end{equation}%
and variance
\begin{equation}
V_{\text{no}}=\frac{1}{2}\Omega =\frac{1}{2}\left( \tau \nu _{\text{th}%
}+V_{\varepsilon }+2\right) .  \label{eq:noise}
\end{equation}

We can then create maximum likelihood estimators for the mean value and
variance of $q_{B_{k}}$ starting from the samples ${B_{k}}_{i}$. In fact, we may write%
\begin{equation}
\widehat{\bar{q}_{B_{k}}}=\frac{N}{m}\sum_{i=1}^{m/N}{B_{k}}_{i},~\widehat{V}%
_{\text{no}_{k}}=\frac{N}{m}\sum_{i=1}^{m/N}\left( {B_{k}}_{i}-\widehat{\bar{%
q}_{B_{k}}}\right) ^{2}.  \label{eq:estimator noise
variance}
\end{equation}%
The mean value and variance of the estimator $\widehat{\bar{q}_{B_{k}}}$ are
given by
\begin{eqnarray}
\mathbb{E}\left( \widehat{\bar{q}_{B_{k}}}\right) &=&\frac{N}{m}%
\sum_{i=1}^{m/N}\mathbb{E}\left( {B_{k}}_{i}\right) =\mathbb{E}(q_{B_{k}}),
\\
\text{Var}\left( \widehat{\bar{q}_{B_{k}}}\right) &=&\frac{N^{2}}{m^{2}}%
\sum_{i=1}^{m/N}\text{Var}\left( {B_{k}}_{i}\right) =\frac{N}{m}V_{\text{no}%
},  \label{varEEE}
\end{eqnarray}%
since ${B_{k}}_{i}$ can be considered to be i.i.d. variables (in a
collective Gaussian attack).

Note that the estimator $\widehat{\bar{q}_{B_{k}}}$ can be replaced by its
expected value $\mathbb{E}(q_{B_{k}})$ due to the fact that its variance in
Eq.~(\ref{varEEE}) vanishes for $m\gg 1$. Thus, we can write the variance
estimator $\widehat{V}_{\text{no}_{k}}$ in Eq.~(\ref{eq:estimator noise
variance}) as
\begin{equation}
\widehat{V}_{\text{no}_{k}}=V_{\text{no}}\frac{N}{m}\sum_{i=1}^{m/N}\left(
\frac{{B_{k}}_{i}-\mathbb{E}(q_{B_{k}})}{\sqrt{V_{\text{no}}}}\right) ^{2}.
\end{equation}%
The term inside the brackets follows a standard normal distribution with
zero mean and unit variance. Therefore, the sum term follows a \PP{non-central} chi-squared
distribution with mean equal to $m/N$ and variance $2m/N$. Consequently, for
the mean and variance of the estimator $\widehat{V}_{\text{no}_{k}}$\ we
obtain
\begin{align}
\mathbb{E}\left[ \widehat{V}_{\text{no}_{k}}\right] =& V_{\text{no}}\frac{N}{%
m}\mathbb{E}\left[\sum_{i=1}^{m/N} \left( \frac{{B_{k}}_{i}-\mathbb{E}%
(q_{B_{k}})}{\sqrt{V_{\text{no}}}}\right) ^{2}\right]  \notag \\
& =V_{\text{no}}, \label{eq:mean noise}\\
\text{Var}\left[ \widehat{V}_{\text{no}_{k}}\right] =& V_{\text{no}%
}^{2}\left( \frac{N}{m}\right) ^{2}\text{Var}\left[\sum_{i=1}^{m/N} \left(
\frac{{B_{k}}_{i}-\mathbb{E}(q_{B_{k}})}{\sqrt{V_{\text{no}}}}\right) ^{2}%
\right]  \notag \\
& =2\frac{N}{m}V_{\text{no}}^{2} \label{eq:variance noise}.
\end{align}

Based on the estimator $\widehat{\bar{q}_{B_{k}}}$ we can build an estimator
for the transmissivity [cf. Eq.~(\ref{eq:arriving signal})]
\begin{equation}\label{eq:tauk estimator}
\hat{\tau}_{k}=2\alpha ^{-2}\cos ^{-2}\left( 2k\pi /N\right) (\widehat{\bar{q%
}_{B_{k}}})^{2}.
\end{equation}%
The estimator $\widehat{\bar{q}_{B_{k}}}$ is the sample mean of ${B_{k}}_{i}$
and as such follows a Gaussian distribution. We then can express Eq.~(\ref{eq:tauk estimator}) with the help of the \PP{non-central} chi-squared variable $\chi_k\equiv\left( \sqrt{\frac{m}{N}}%
\frac{\widehat{\bar{q}_{B_{k}}}}{\sqrt{V_{\text{no}}}}\right) ^{2}$ as follows:
\begin{equation}
\hat{\tau}_{k}=2\frac{V_{\text{no}}}{[\alpha \cos \left( 2k\pi /N\right)
]^{2}}\frac{N}{m}\left( \sqrt{\frac{m}{N}}\frac{\widehat{\bar{q}_{B_{k}}}}{%
\sqrt{V_{\text{no}}}}\right) ^{2}.
\end{equation}%
Because $\chi_k$ has
mean value $1+\frac{m}{N}\frac{\tau \lbrack \alpha \cos \left( 2k\pi /N\right)
]^{2}}{2V_{\text{no}}}$ and variance $2\left( 1+2\frac{m}{N}\frac{\tau
\lbrack \alpha \cos 2k\pi /N]^{2}}{2V_{\text{no}}}\right) $,
the estimator of the transmissivity has mean and variance equal to
\begin{align}
\mathbb{E}\left( \hat{\tau}_{k}\right) & =\frac{2V_{\text{no}}N}{m\alpha
^{2}\cos ^{2}\left( 2k\pi /N\right) }  \notag \\
& \times \left( 1+\frac{m}{N}\frac{\tau \lbrack \alpha \cos \left( 2k\pi
/N\right) ]^{2}}{2V_{\text{no}}}\right)  \notag \\
& =\tau +\mathcal{O}(1/m), \\
\text{Var}\left( \hat{\tau}_{k}\right) & :=\sigma _{k}^{2}=\left( \frac{2V_{%
\text{no}}N}{m\alpha ^{2}\cos ^{2}\left( 2k\pi /N\right) }\right) ^{2}
\notag \\
& \times 2\left( 1+2\frac{m}{N}\frac{\tau \lbrack \alpha \cos \left( 2k\pi
/N\right) ]^{2}}{2V_{\text{no}}}\right)  \notag \\
& =8\tau \frac{N}{m}\frac{V_{\text{no}}}{\alpha ^{2}\cos ^{2}\left( 2k\pi
/N\right) }+\mathcal{O}(1/m^{2}).
\end{align}

Since there will be other estimators corresponding to the other values of
Alice's encoding $k$, we can create an optimal linear combination of them
with variance~\cite{Usenko_cpe}
\begin{align}
\sigma _{q}^{2}& =\left[ \sum_{k=0}^{N-1}({\sigma _{k}^{2})}^{-1}\right]
^{-1}  \notag \\
& =8\tau \frac{N}{m}\frac{V_{\text{no}}}{\alpha ^{2}}\left[
\sum_{k=0}^{N-1}\cos \left( 2k\pi /N\right) \right] ^{-1}  \notag \\
& =\tau \frac{16}{m}\frac{V_{\text{no}}}{\alpha ^{2}}.
\end{align}%

So far, we used only samples from the $q$-quadrature of Bob's outcomes.
Similar relations will hold for the $p$-quadrature. Combining all the
available $q$- and\ $p$-samples, the optimal linear estimator $\hat{\tau}$
of the transmissivity will have
\begin{equation}
\mathbb{E}(\hat{\tau})=\tau,~~\text{Var}(\hat{\tau}):=\sigma ^{2}=\tau \frac{8}{m}\frac{V_{\text{no}}}{\alpha ^{2}}\label{eq: mean,varaince tau}.
\end{equation}%
In fact, for large $m$, we can approximate all the $2N$ estimators $\hat{\tau%
}_{k}$ to have Gaussian distributions with the same mean and variance $%
\sigma _{p}^{2}=\sigma _{q}^{2}$. As a result, the global estimator $\hat{%
\tau}$ is a Gaussian variable with the same mean $\tau$ and variance equal to $%
\sigma ^{2}$. Now, assuming an error $\epsilon _{PE}=10^{-10}$
for channel parameter estimation (PE), we have to consider a $6.5$ standard deviation
interval for $\tau$. This means that the worst-case value for the transmissivity
is equal to
\begin{equation}
\tau ^{\epsilon _{\text{PE}}}=\tau -6.5\sqrt{\tau \frac{8}{m}\frac{V_{\text{%
no}}}{\alpha ^{2}}}.
\end{equation}

Starting from $\widehat{V}_{\text{no}_{k}}$ we may also define an estimator
for the excess noise. Solving Eq.~(\ref{eq:noise}) with respect to $%
V_{\varepsilon }$, we obtain
\begin{equation}
\widehat{V}_{\varepsilon _{k}}=2\widehat{V}_{\text{no}_{k}}-\hat{\tau}\nu _{%
\text{th}}-2.
\end{equation}%
Then the mean and variance of this estimator are given by
\begin{align}
\mathbb{E}\left[ \widehat{V}_{\varepsilon _{k}}\right] & =2V_{\text{no}%
}-\tau \nu _{\text{th}}-2, \\
\text{Var}\left[ \widehat{V}_{\varepsilon _{k}}\right] & :=s_{k}^{2}=8\frac{N%
}{m}V_{\text{no}}^{2}+\sigma ^{2}\nu^2 _{\text{th}},
\end{align}%
where we used Eqs.~(\ref{eq:mean noise}),~(\ref{eq:variance noise}) and~(\ref{eq: mean,varaince 
tau}).
The variance of the optimal linear combination $\widehat{V}_{\varepsilon }$
of all the estimators $\widehat{V}_{\varepsilon _{k}}$ (also considering the
$p$-quadrature) is given by
\begin{equation}
s^{2}=\frac{4V^2_{\text{no}}}{m}+\frac{\sigma ^{2}\nu^2 _{\text{th}}}{2N}.
\end{equation}%

Based on the assumption of large $m$, we approximate the distribution of
each $\widehat{V}_{\text{no}_{k}}$ to be Gaussian. As a result, the
distribution of $\widehat{V}_{\varepsilon }$ is Gaussian with the same mean
and variance given by $s^{2}$ above. Assuming an error
$\epsilon _{PE}=10^{-10}$, we obtain the $6.5$ confidence intervals
for $\widehat{V}_{\varepsilon }$. Therefore, the worst-case value is give by%
\begin{equation}
V_{\varepsilon }^{\epsilon _{\text{PE}}}=V_{\varepsilon }+6.5\sqrt{\frac{4 V_{%
\text{no}}^{2}}{m}+\frac{\sigma ^{2}\nu^2 _{\text{th}}}{2N}}.
\end{equation}

Using the worst-case values $\tau ^{\epsilon _{\text{PE}}}$ and $%
V_{\varepsilon }^{\epsilon _{\text{PE}}}$, we can write a finite-size
expression of the key rate $R=R(\tau ,V_{\varepsilon })$ of Eq.~(\ref%
{eq:asymptotic rate 2}) which accounts for the imperfect parameter
estimation and the reduced number of signals. This is given by replacing
\begin{equation}
R(\tau ,V_{\varepsilon })\rightarrow \frac{n}{M}R(\tau ^{\epsilon _{\text{PE%
}}},V_{\varepsilon }^{\epsilon _{\text{PE}}}):=\frac{n}{M}R_{\epsilon _{%
\text{PE}}}.  \label{RCPE}
\end{equation}

\section{Composable security under collective attacks\label{composableSEC}}
\PP{The following study for the composable security is based on various ingredients~\cite{Cosmo_comp,Portmann,free_space,TomaThesis,AEP,compLeverier1,TomaRenner}. More precisely, it follows the procedure formulated in Ref.~\cite{free_space} which is here specified and applied to a discrete alphabet.}

After the parties exchange $n$ signal states and apply error correction (EC), they share
a state $\tilde{\rho}^n$ from which, according to the leftover hash lemma, they can
extract $s_n$ bits of uniform randomness, or in other words secret key bits. This
number of bits is bounded according to the following relation~\cite{TomaRenner,TomaThesis}:
\begin{equation}
s_{n}\geq H_{\text{min}}^{\epsilon_{\text{s}}}(l^{n}|E^{n})_{\tilde{\rho}^{n}%
}+2\log_{2}\sqrt{2}\epsilon_{\text{h}}-\mathrm{leak}_{n,\text{EC}}(n,\epsilon_\text{cor}). \label{eeq2}%
\end{equation}
Here, $H_{\text{min}}^{\epsilon_{\text{s}}}(l^{n}|E^{n})$ is the smooth min-entropy of Bob's variable $l$ conditioned on Eve's
systems $E$, and $\mathrm{leak}_{\text{EC}}(n,\epsilon_\text{cor})$ is the classical information exchanged by the parties for EC (stored by Eve in her register). 

The uniform randomness $\epsilon_\PP{\text{h}}$ and smoothing  $\epsilon_\PP{\text{s}}$ parameters  define the secrecy of the protocol $
\epsilon_\text{sec}=\epsilon_\PP{\text{h}}+\epsilon_\PP{\text{s}}$ which, along with the EC parameter $\epsilon_\text{cor}$, defines the security parameter $\epsilon_\text{tot}=
\epsilon_\text{cor}+\epsilon_\text{sec}$. The latter  bounds the trace distance $D$ of the state $\bar{\rho}^n$ (after privacy amplification)
from the ideal output state $\rho_\text{id}$ of a QKD protocol, i.e., a classical-quantum state where the uniformly distributed classical
registers of Alice and Bob are uncorrelated from Eve's systems~\cite{Portmann}.
Each of the epsilon parameters introduced above can be considered to be very small. We take them of the order of $10^{-10}$.

Equation~(\ref{eeq2}) can be further simplified so as to be connected with the asymptotic secret key rate. In 
fact, we can further bound the smooth min entropy calculated in terms of $\tilde{\rho}^n$ with the 
smooth min entropy
of the state before EC $\rho^{\otimes n}$, which is in a tensor-product form due to the fact that we  
assumed a collective attack. More precisely, \PP{we exploit the following inequality~\cite{free_space}}:
\begin{align}
H_{\text{min}}^{\epsilon_{\text{s}}}(l^{n}|E^{n})_{\tilde{\rho}^{n}}  &  \geq
H_{\text{min}}^{\PP{p\epsilon_\text{s}^2/3}}(l^{n}|E^{n})_{\rho^{\otimes
n}}\nonumber \\
&  +\log_{2}\left[p\left(  1-\PP{\epsilon^2_{\text{s}}/3} \right)  \right].\label{eeq3}
\end{align}
Here $p$ is the probability of successful EC, i.e., the probability that the protocol is not aborted after Alice and Bob  compared hashes of their sequences~\cite{Stefano_rev2}. The value of $1-p$ is given by the experimental frame error rate~\cite{LeoEXP}. Note that, even if the protocol does not abort (because the hashes are the same), Alice's and Bob's sequences are identical up to an error probability $\epsilon_\text{cor}$.

The  replacement in Eq.~(\ref{eeq3}) allows us to use the asymptotic equipartition property (AEP) theorem~\cite{AEP} so as to reduce the conditional smooth-min entropy  of the tensor-product form $\rho^{\otimes n}$ to the conditional von Neumann entropy $S(l|E)_{\rho}$ of the single copy $\rho$. In particular, one may write the following~\cite{TomaThesis}:
\begin{align}
H_{\text{min}}^{\PP{p \epsilon_{\text{s}}^2/3}}(l^{n}|E^{n})_{\rho^{\otimes
n}}  &  \geq n S(l|E)_{\rho}\nonumber\\
&  -\sqrt{n}\Delta_{\text{AEP}}\left( \PP{p\epsilon_{\text{s}}^2/3},|\mathcal{L}|\right)  , \label{eeq4}
\end{align}
where
\begin{equation}
\Delta_{\text{AEP}}(\epsilon_{\text{s}},|\mathcal{L}|):=4\log_{2}\left(  2\sqrt
{|\mathcal{L}|}+1\right)  \sqrt{\log(2/\epsilon_{\text{s}}^{2})}. \label{AEPd}%
\end{equation}
The parameter $|\mathcal{L}|$ is the cardinality of Bob's outcome (alphabet) and, in our case, it is equal to $N$. 

Replacing Eqs.~(\ref{eeq3}) and~(\ref{eeq4}) in  Eq.~(\ref{eeq2}), one obtains the following bound for the number of secret bits:
\begin{gather}
s_{n}\geq nS(l|E)_{\rho}-\sqrt{n}\Delta_{\text{AEP}}\left(\PP{p \epsilon_{\text{s}}^2/3},N\right)  \nonumber\\
+\log_{2}[p\left(  1-\PP{\epsilon^2_{\text{s}}/3}\right)  ]+2\log_{2}\sqrt{2}\epsilon
_{\text{h}}-\mathrm{leak}_{\text{EC}}(n,\epsilon_\text{cor}).\label{form}%
\end{gather}
To further simplify the bound above, consider the definition of quantum mutual information between two systems $Q$ and $E$ in terms of the (conditional) von Neumann entropy
\begin{equation}
I(Q:E)=S(Q)-S(Q|E).
\end{equation}
When $Q$ is a classical system described by a variable $l$, $I(l:E)$ takes the form of the Holevo information $\chi(E:l)$ and the von Neumann entropy simplifies to the Shannon entropy $H(l)$. Thus we can write
\begin{equation}
S(l|E)_{\rho}=H(l)_{\rho}-\chi(E:l)_{\rho}.\label{ff1}%
\end{equation}

Moreover, let us set the quantity
\begin{equation}
H(l)_{\rho}-n^{-1}\mathrm{leak}_{\text{EC}}(n,\epsilon_{\text{cor}}):=\xi I(k:l)_{\rho},\label{ff2}%
\end{equation}
where $I(k:l)$ is the classical mutual information between Alice's and Bob's variables and $\xi \in [0,1]$ defines the reconciliation efficiency~\cite{receff}. As a result, the asymptotic secret key rate of Eq.~(\ref{eq:asymptotic rate 2}) appears if we make the previous replacements in Eq.~(\ref{form}) obtaining
\begin{align}
s_{n} &  \geq n R_{\rho}-\sqrt{n}\Delta_{\text{AEP}}\left( \PP{ p \epsilon_{\text{s}}^2/3},N\right)  \nonumber\\
&  +\log_{2}[p\left(  1-\PP{\epsilon^2_{\text{s}}/3}\right)  ]+2\log_{2}%
\sqrt{2}\epsilon_{\text{h}}.\label{snprima}%
\end{align}
\begin{figure}
\vspace{-0.1cm}
\includegraphics[width=0.42\textwidth]{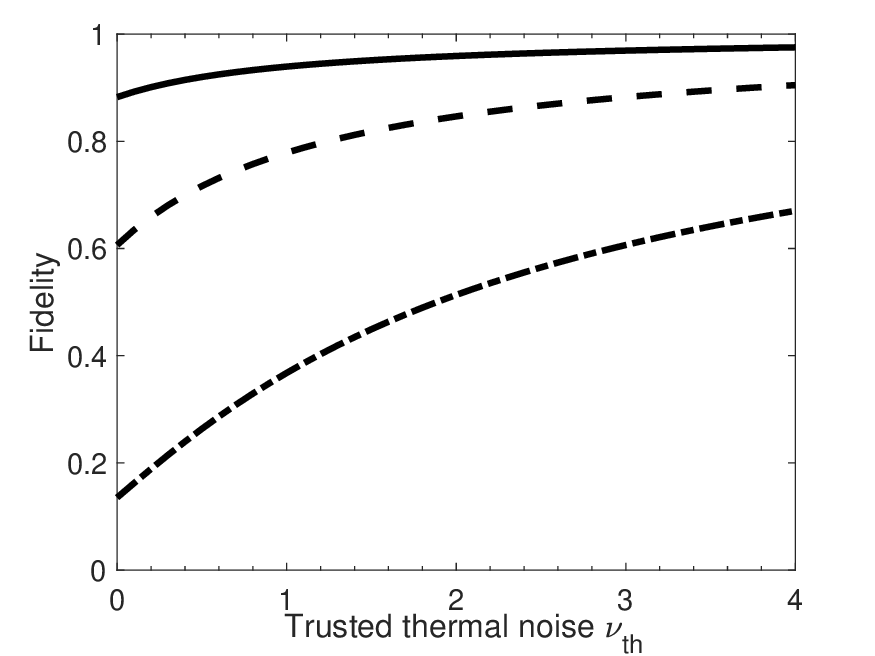}
\vspace{-0.2cm}
\caption{\label{fig:orthogonality} The fidelity between two signal states ($N=2$) for $k=0$ and $k=1$ versus the thermal preparation noise $\nu_\text{th}$. We include plots for different amplitudes $\alpha=0.5$ (solid line), $\alpha=1$ (dashed line) and $\alpha=2$ (dashed-dotted line). As the thermal noise increases,  the fidelity between the two states arrives at a saturation point close to $1$. The smaller the value of $\alpha$ the faster this saturation happens.}
\end{figure}
Finally, let us account for the PE in the bound above. This means that we need to write Eq.~(\ref{snprima}) considering the worst-case scenario state $\rho^{M-m}_{\epsilon_\text{PE}}$,  where the channel parameters $\tau$ and $V_\varepsilon$ are bounded by  $\tau^{\epsilon_\text{PE}}$ and $V_\varepsilon^{\epsilon_\text{PE}}$, and also accounting for the fact that we sacrificed $m$ out of $M$ signal states. Therefore, we obtain
\begin{align}
s_{M-m} &  \geq (M-m) R_{\epsilon_\text{PE}}-\sqrt{M-m}\Delta_{\text{AEP}}\left( \PP{ p \epsilon_{\text{s}}^2/3},N\right)  \nonumber\\
&  +\log_{2}[p\left(  1-\PP{\epsilon^2_{\text{s}}/3}\right)  ]+2\log_{2}%
\sqrt{2}\epsilon_{\text{h}},\label{snprimaPE}%
\end{align}
where  $R_{\epsilon_\text{PE}}$ is the finite-size rate of Eq.~(\ref{RCPE}).
This is true only with probability $1-\epsilon_\text{PE}$ since there is a non-zero probability $\epsilon_\text{PE}$ that the actual values of the channel parameters are not bounded by $\tau^{\epsilon_\text{PE}}$ and $V_\varepsilon^{\epsilon_\text{PE}}$. Dividing Eq.~(\ref{snprimaPE}) by the total number $M$ of signal states, multiplying by the EC success probability $p$, and setting $r=\frac{m}{M}$, we obtain the secret key rate
\begin{align}
R_{M,r} \geq&(1-r)p\left[R_{\epsilon_\text{PE}}-\frac{1}{\sqrt{(1-r)M}}\Delta_{\text{AEP}}\left(  \PP{p \epsilon_{\text{s}}^2/3},N\right) \right.\nonumber\\
&  + \frac{\log_{2}[p\left(  1-\PP{\epsilon^2_{\text{s}}/3})\right]+2\log_{2}\sqrt{2}\epsilon_{\text{h}}}{(1-r)M}\Bigg],\label{sckeee}%
\end{align}
which is valid up to an overall $\epsilon_\text{tot}=\epsilon_{\text{cor}}+\epsilon_{\text{s}}+\epsilon_{\text{h}}+p \epsilon_{\text{PE}}$. \PP{See Ref.~\cite{free_space} for corresponding analytical
formulas but in the setting of Gaussian-modulated protocols.}


In Fig.~\ref{fig:plot_correct}, we plot the composable key rate for the protocol with two states ($N=2$) versus the attenuation in dB for $r=0.01$, $M=10^{12}$ (blue dashed line) and  $M=10^{9}$ (blue solid line). We assume excess noise $\varepsilon=0.01$ and set the security parameters to $\epsilon_\PP{\text{s}}=\epsilon_\PP{\text{h}}=\epsilon_{\text{PE}}=10^{-10}$. We assume that the reconciliation efficiency parameter is $\xi=0.99$ and the EC success probability is $p=0.9$. (Note that, in our analysis the EC error $\epsilon_{\text{cor}}$ is contained in $\xi$.) In Fig.~\ref{fig:plot_correctN=3}, we plot the secret key rate for $N=3$, both in the asymptotic (black line) and the composable cases (blue lines) for channel excess noise $\varepsilon=0.01$. As expected, the performance of the protocol is dependent on the number $M$ of signals.

\begin{figure}[h]
\vspace{-0.2cm}
\includegraphics[width=0.42\textwidth]{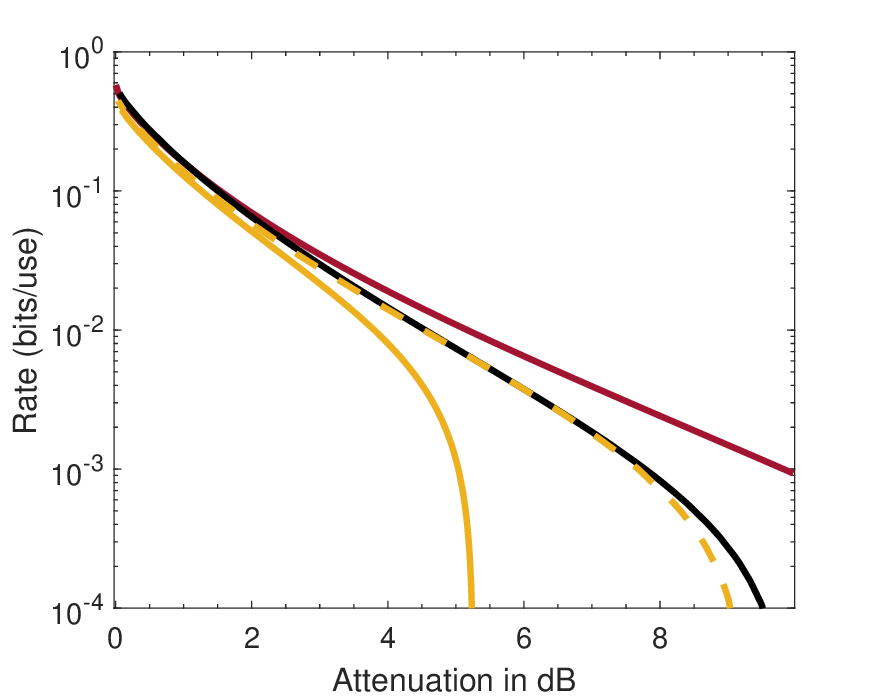}
\caption{\label{fig:thermalrate_thermalchannel}The secret key rate for $N=2$ and $\nu_\text{th}=0.1$ versus the attenuation in dB. We  assumed $\alpha=2$ and excess noise $\varepsilon=0.01$. We include the asymptotic case  with $\xi=1$ (red solid line) and the composable case, for which we assume $\epsilon_\PP{\text{s}}=\epsilon_\PP{\text{h}}=10^{-10}$, $\epsilon_\PP{\text{PE}}=10^{-10}$, $p=0.9$, $\xi=0.99$ and $r=0.01$, for  $M=10^{12}$ (yellow dashed line) and  $M=10^{9}$ (yellow solid line). For comparison, we also plot the secret key rate assuming coherent states ($\nu_{\text{th}}=0$, black solid line). We observe an advantage when we use preparation trusted noise (compare red and black lines) that can be exploited to mitigate the decrease of the rate in the finite-size regime (similar performance of yellow dashed line and black solid line). All the lines have a truncation accuracy of $14$ Fock-basis states.}
\end{figure}

As we can see in Fig.~\ref{fig:orthogonality}, increasing preparation (trusted) thermal noise $\nu_\text{th}$~\cite{usenkotrusted}, the fidelity of the signal states increases, making them more difficult to distinguish, resulting in a better secret key rate performance. In more detail, we observe that the fidelity (computed according to Ref.~\cite{BanchiPRL}) reaches a saturation point faster when $\alpha$ is smaller. Furthermore, in this point the fidelity becomes closer to $1$ as $\alpha$ gets smaller. Taking into consideration the channel propagation, this  leads to a configuration where Bob's states may have almost the initial fidelity,
while the fidelity of Eve's states may be at the saturation point. This can happen for example for transmissivities that are close to $1$. An additional optimal value of the thermal preparation noise can boost this effect for other transmissivities. In fact, we can observe this in Fig.~\ref{fig:thermalrate_thermalchannel}, where we consider excess noise $\varepsilon=0.01$ and preparation noise $\nu_{\text{th}}=0.1$, i.e., Alice sending thermal states. We observe an advantage for the secret key rate when we use preparation noise that compensates the rate degradation due to the finite-size effects.

\section{Conclusion and discussion\label{ConclusionSEC}}
In this work, we  study the finite-size composable security of a
discrete-alphabet CV-QKD protocol under the assumption of collective
Gaussian attacks. This assumption is realistic because the standard model of
loss and noise in optical quantum communications is the memoryless
thermal-loss channel, which is dilated into a collective entangling cloner
attack, i.e., a specific type of collective Gaussian attack~\cite{collectiveG}. Our analysis extends previous asymptotic analyses~\cite{disc9,disc10} to the
finite-size and composable regime, but simultaneously pays the price to be
restricted to collective Gaussian attacks. Removing this assumption is the subject of
future investigations.

Since our analysis applies not only to displaced coherent states but also to
displaced thermal states, it can be useful for studying the security of
phase-encoded protocols at frequencies lower than the optical. Moreover,
the use of displaced thermal states  can increase the difficulty in
distinguishing the signal states with a beneficial effect for the secret
key rate. It is also worth stressing that our derivation, described for phase-encoded signals, can immediately be extended to any constellation of
displaced Gaussian states (e.g., coherent, thermal or squeezed). The most crucial part is the finite-size rate $R_{\epsilon _{\text{PE}}}$ which can always be estimated, under the assumption of
collective Gaussian attacks, by using maximum likelihood estimators
and their confidence intervals, i.e., adopting simple variations of the technique in Sec.~\ref{sec:CPE}.
In this way, the finite-size rate 
$R_{\epsilon _{\text{PE}}}$ can always be expressed in terms of the asymptotic key rate $R$ of the specific protocol under consideration via the
transformation in Eq.~(\ref{RCPE}). 

Note that, for Gaussian-modulated coherent-state protocols, one can apply a Gaussian de Finetti reduction~\cite{Lev2017} that enables one to extend the composable security to general coherent attacks. However, this technique does not seem to be applicable to coherent-state protocols with discrete, finite alphabets.
Finally, also note that our study involves calculations based on a suitable truncation of the Fock space. The computational cost associated with these calculations is discussed in the Appendix.

\smallskip

\textbf{Acknowledgements}.~~The authors would like to thank Q. Liao and C. Ottaviani for feedback. This work was sponsored by the
the European Union
via \textquotedblleft Continuous Variable Quantum
Communications\textquotedblright\ (CiViQ, Grant agreement No 820466) and the EPSRC via the
Quantum Communications Hub (Grant No. EP/T001011/1). The computational part of this study was carried out on the Viking Cluster of the University of York.
\appendix

\section{Computational cost\label{app:computational cost}}
\begin{figure}[t]
\vspace{-0.1cm}
\includegraphics[width=0.35\textwidth]{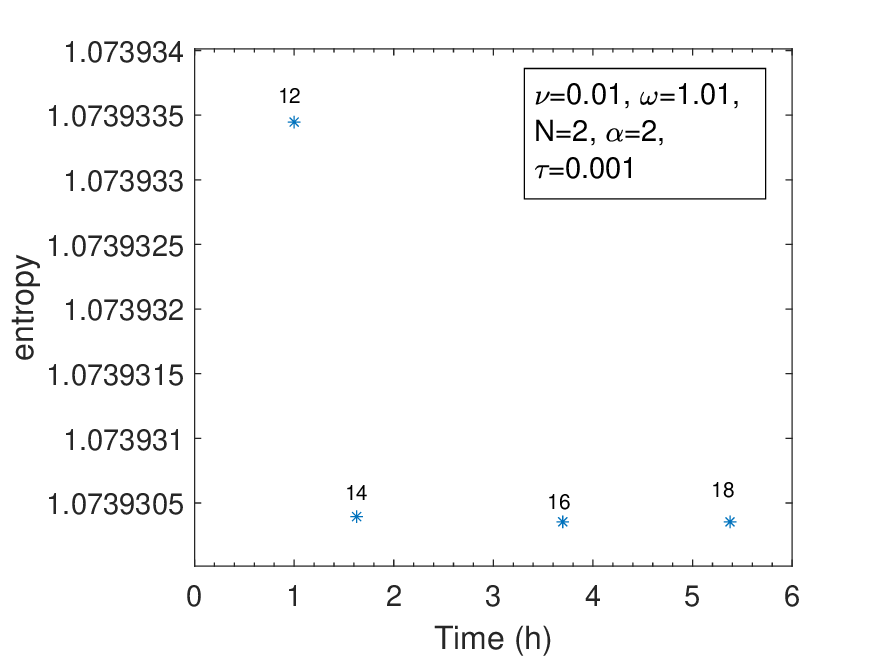}
\vspace{-0.2cm}
\caption{\label{fig:case_six} Entropy convergence versus computational time. Every point has been plotted with a different truncation number in the Fock basis which is indicated on top of the point. As this number is getting larger the entropy is converging to its actual value. For keeping the sixth significant digit, we need to truncate at $14$ states, which needs about $1.5 h$. For this plot, we used preparation noise $\nu=0.01$, channel thermal noise $\omega=1.01$, number of ensemble states $N=2$, amplitude $\alpha= 2$ and channel transmissivity $\tau=0.001$. }
\end{figure}

\begin{figure}[h]
\vspace{-0.1cm}
\includegraphics[width=0.35\textwidth]{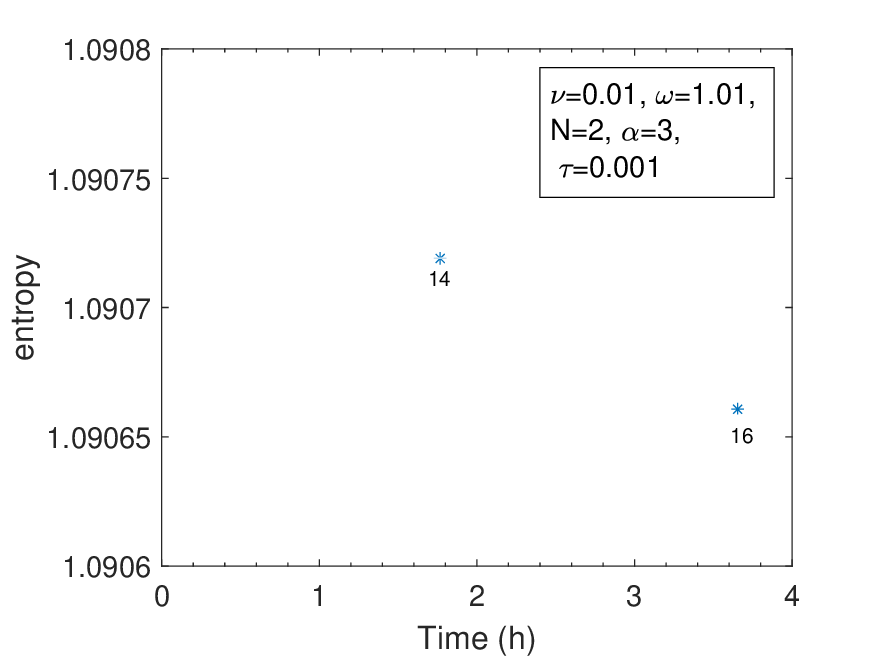}
\vspace{-0.2cm}
\caption{\label{fig:case_seven} Entropy convergence versus computational time. Here we present the case of Fig.~\ref{fig:case_six} but for $\alpha=3$. For keeping the fourth significant digit we need to truncate at $16$ states or more. }
\end{figure}

\begin{figure}[t]
\vspace{-0.1cm}
\includegraphics[width=0.35\textwidth]{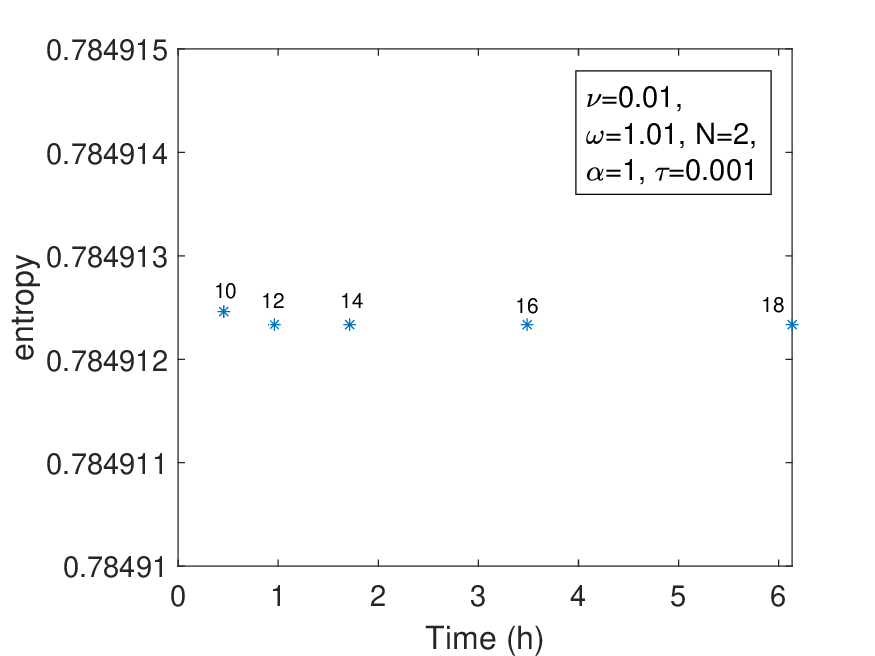}
\vspace{-0.2cm}
\caption{\label{fig:case_five} Entropy convergence versus computational time. For completeness, we  present the case of Fig.~\ref{fig:case_six} but for $\alpha=1$, where $10$ to $12$ states could be enough if we want to keep the fifth significant digit.}
\end{figure}

\begin{figure}[t]
\vspace{-0.1cm}
\includegraphics[width=0.35\textwidth]{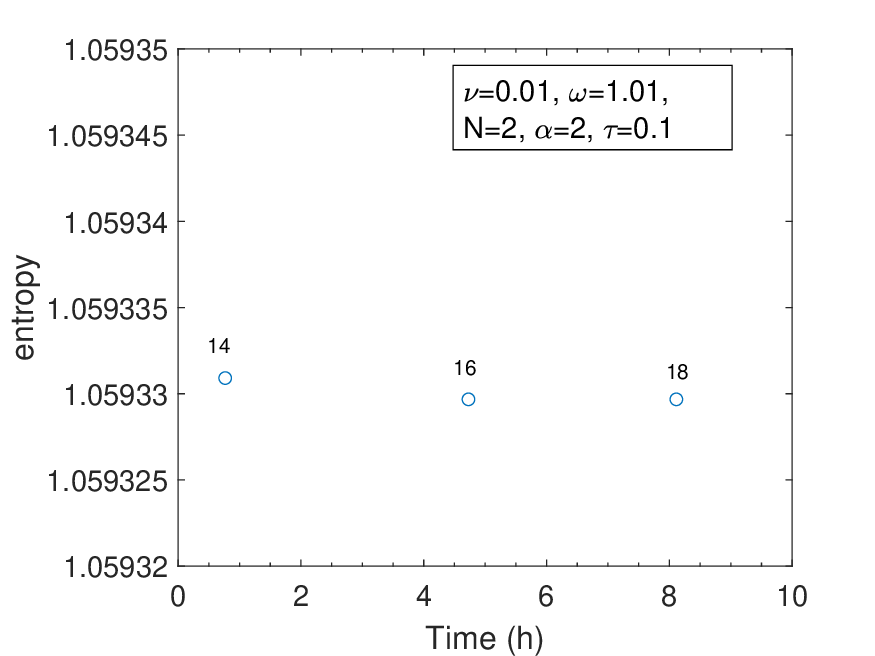}
\vspace{-0.2cm}
\caption{\label{fig:tau=0.1} Entropy convergence versus computational time. We  present the case of Fig.~\ref{fig:case_six} but for $\tau=0.1$. Here we need at least $14$ states for keeping the  fourth significant digit.}
\end{figure}

\begin{figure}[t]
\vspace{-0.1cm}
\includegraphics[width=0.35\textwidth]{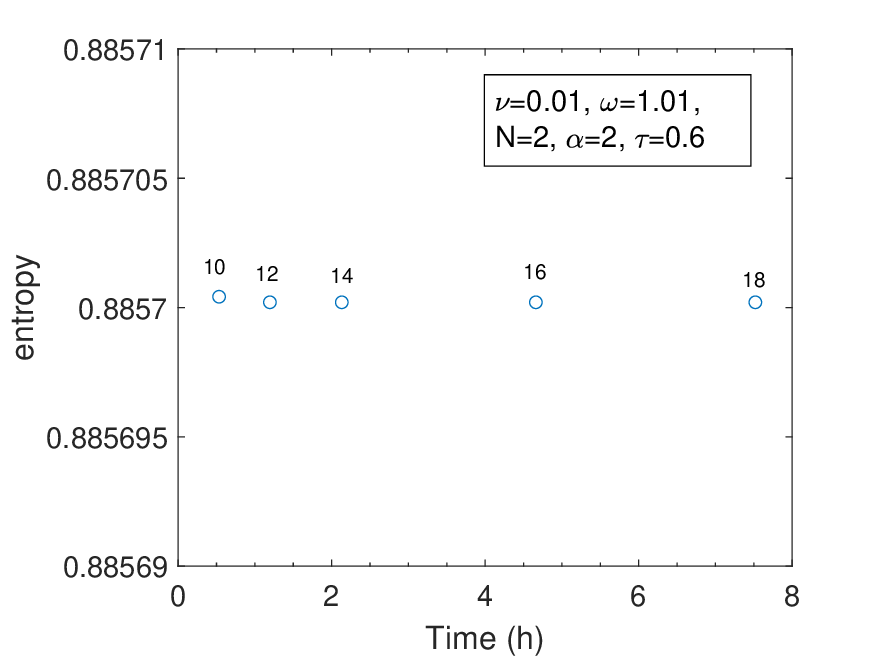}
\vspace{-0.2cm}
\caption{\label{fig:case_two} Entropy convergence versus computational time. We present the case of Fig.~\ref{fig:case_six} but for $\tau=0.6$. We observe that as the transmissivity increases, the convergence becomes faster.}
\end{figure}

\begin{figure}[t]
\vspace{-0.1cm}
\includegraphics[width=0.35\textwidth]{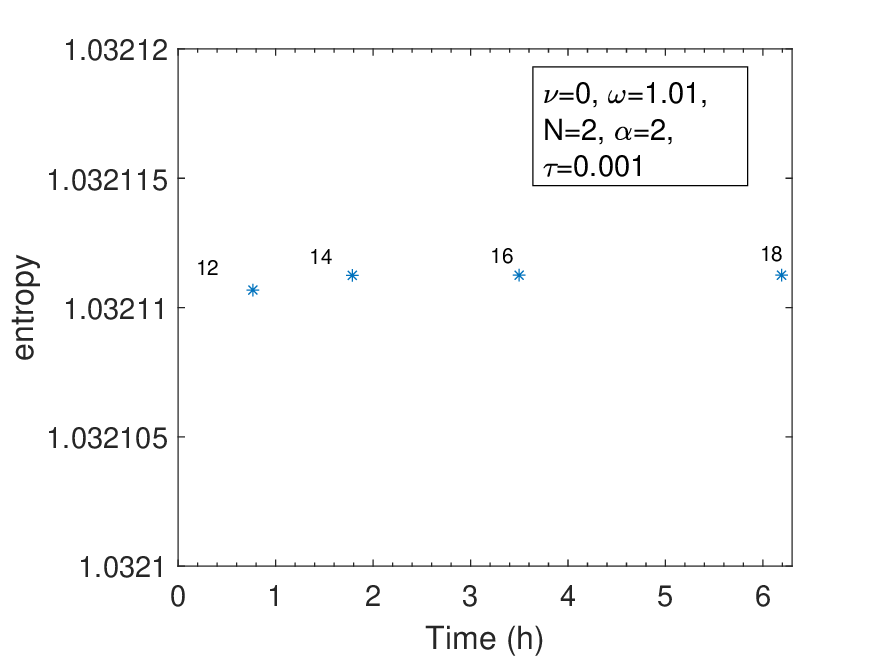}
\vspace{-0.2cm}
\caption{\label{fig:nu0}  Entropy convergence versus computational time. For completeness, we present the case of Fig.~\ref{fig:case_six} but for $\nu=0$.}
\end{figure}

Here we present some results on the computational cost associated with the calculation of the secret key rate. More precisely, we focus on the time required for the convergence of the entropy of Eve's average state as the number of Fock states increases. This functional is simpler to examine and provides a good estimate of the time needed for the full key rate. Note that, as the transmissivity decreases, i.e., the distance increases, the secret key becomes lower and lower; for this reason, we need more significant digits to approximate it. Correspondingly, this demands extra significant digits for the calculation of the entropy. In Figs.~\ref{fig:case_six}, \ref{fig:case_seven} and   \ref{fig:case_five} we compare this convergence for different values of $\alpha=1,2,3$ while we set: preparation noise $\nu=0.01$, channel noise $\omega=1.01$, and transmissivity $\tau=0.001$.
From Figs.~\ref{fig:case_six},~\ref{fig:tau=0.1} and~\ref{fig:case_two}, we may also compare the performances with different values of the transmissivity $\tau=0.001,0.1,0.6$. In Fig.~\ref{fig:nu0}, we present the corresponding case for zero preparation noise. Calculations were performed using \textsc{MatLab} 2018 on a core of a 20-core  2.0 GHz  Intel Xeon 6138  of the Viking cluster~\cite{VIKINGCLUSTER}.

\begin{thebibliography}{99}
\bibitem{Note} This state is calculated by using \cite[Eq.~(A7)]{Stefano_rev2}
replacing its mean and CM from Eq.~(\ref{eq:mean heterodyne}) and Eq.~(\ref{eq:CM heterodyne}) and expressing the quadrature operators in terms of the Fock basis considering the appropriate truncation.
\bibitem{receff}
The analytical dependence of leak$_\text{EC}(n,\epsilon_\text{cor})$ on $\epsilon_\text{cor}$ is determined in an experimental scenario after the choice of the linear EC code. At that point, one can more precisely relate $\xi$ to $\epsilon_\text{cor}$ through Eq.~(\ref{ff2}). From typical experiments for Gaussian protocols, it is known that $\xi=0.95$~\cite{Jouguetexp} and $\xi=0.98$~\cite{Milicevicrec,receffvalue} is achievable. In the case of  discrete-alphabet encoding the EC is very efficient. This is why we chose a reconciliation efficiency equal to $\xi=0.99$.
\end{thebibliography}
\end{document}